\documentclass[onecolumn,showpacs,preprintnumbers,amsmath,amssymb,aps]{revtex4}
\usepackage{bm}
\usepackage{color}

\usepackage{graphicx,subfig,epsfig,epsf,color}
\usepackage[utf8x]{inputenc}
\usepackage{amssymb,amsmath,appendix}
\usepackage{slashed}
\usepackage{array}
\usepackage{hyperref}

\begin{document}
\title{Variation of delta baryon mass and hybrid star properties in static and rotating conditions}
\author{Debashree Sen}
\email[]{debashreesen88@gmail.com}
\affiliation{Department of Physical Sciences\\
Indian Institute of Science Education and Research Berhampur,\\
Transit Campus, Government ITI, 760010 Berhampur, Odisha, India}

\date{\today}




\begin{abstract}

The possible conditions for hadron-quark phase transition in hybrid star cores are investigated in the present work. For the hadronic matter part the effective chiral model is adopted. Exotic baryonic degrees like hyperons and the delta baryons are also taken into account. As $\Delta$s posses Breit-Wigner mass distribution ($1232 \pm 120$ MeV), the hadronic equation of state is obtained by varying the mass of the delta baryons in this range. For the quark phase the MIT bag model is chosen with repulsive effects of the unpaired quarks. Phase transition is achieved using Gibbs construction and the gross properties of the resultant hybrid star are calculated in both static and rotating conditions and compared with the various constraints on them from different observational and empirical perspectives. The work presents a thorough study of the phase transition properties like the critical density of appearance of quarks, the density range for the persistence of the mixed phase and the population of different hadrons and quarks in hybrid star matter. The hybrid star properties, calculated in both static and rotating conditions, are found to be consistent with the bounds on them from different perspectives.


\end{abstract}




\maketitle


\section{Introduction}
\label{intro}

 The composition of neutron star (NS) matter (NSM) at high density (($5-10)\rho_0$; $\rho_0\approx0.16$ fm$^{-3}$ being the normal nuclear matter density) is one of the most interesting and active research areas of NS physics. Nuclear matter properties are mostly well-examined in the vicinity of saturation density $\rho_0$. However, at high density relevant to NS cores, the properties of matter and interactions are still inconclusive from experimental perspectives. Therefore at present the composition of NSM still remain uncertain. The theoretical calculations of the structural properties of NSs are largely dependent on the equation of state (EoS) of NS, which in turn is determined by the composition and the interactions considered. Theoretical predictions have suggested the possibilities of formation of exotic baryons like the hyperons \cite{Glendenning,Miyatsu2012,Bednarek2012,Weissenborn2012,Weissenborn2014, Agrawal2012,Lopes,Oertel,Colucci,Dalen,Lim,Rabhi2012,Sen2,Sen3,Baldo,Vidana2, Katayama,Yamamoto}, $\Delta$ baryons \cite{Boguta1982,Glendenning,Cai,Sun,Kolomeitsev,Maslov,Zhu,Drago2014,Li2018,Sen, Sen4} at high density when the nucleon chemical potential matches with the rest masses of these heavier baryons. Although \cite{Glendenning,Glen85} predicted that formation of $\Delta$s are not favored in NSM, recent studies based on both relativistic mean field (RMF) \cite{Drago2014,Cai,Kolomeitsev,Sen,Sen4} and microscopic \cite{Zhu,Li2019,Logoteta16} approaches have not only shown early onset of $\Delta$s but also predicted that they may populate NSM considerably and bring significant changes to structural properties of the NSs especially the radius and compactness \cite{Drago2014,Cai,Zhu,Li2019,Sen,Sen4}. 
 
 On the other hand, it has been suggested from the QCD phase diagram that at very high temperature or density, hadronic matter is prone to undergo phase transition to form deconfined quark matter composed of u, d and s quarks \cite{Glendenning,Weissenborn2011,Ozel2010,Klahn,Bonanno,Lastowiecki,Drago2016,Drago2016(2), Zdunik,Masuda,Wu,Sen,Sen2}. Thus the possibility of such transition at high density relevant to NS cores, thereby forming hybrid stars (HSs), is of great current interest. Recently, in the context of binary neutron star merger (BNSM), works like \cite{Alford19} have suggested that hadron-quark phase transition may be possible in NS cores. 
 
 However, such exotic degrees of freedom are known to soften the EoS that dictates the gross NS properties. At present, certain bounds obtained on the different properties of neutron/compact stars from various perspectives constrain the EoS to some extent. The last decade was extremely successful in this regard. The discovery of the most massive pulsars like PSR J0348+0432 \cite{Ant} and PSR J0740+6620 \cite{Cromartie} that have put upper bounds on the gravitational mass. Also, with the phenomenal detection of gravitational waves (GW170817) from BNSM by LIGO-Virgo collaboration, stringent bounds on dimensionless tidal deformability ($\Lambda_{1.4}$) and radius ($R_{1.4}$) of a 1.4$M_{\odot}$ NS are obtained \cite{Abbott,Fattoyev,Most}. It also opened up new windows to constraint indirectly and co-relate several other NS properties like the symmetry energy \cite{Tong2020}, speed of sound in NSM \cite{Kanakis-Pegios,NaZhang,cs3,Reed,Marczenko} and many more in terms of $\Lambda_{1.4}$ and $R_{1.4}$. Moreover, very recently, constraints on the $M-R$ relation have been obtained from PSR J0030+0451 in the NICER experiment \cite{Miller}. Apart from the gravitational mass and radius, the maximum bounds on surface redshift ($Z_s$) are established from the source spectrum analysis of 1E 1207.4-5209 \cite{Sanwal} and RX J0720.4-3125 \cite{Hambaryan}. As NSs are mostly observed as pulsar or rotating NSs, therefore it becomes imperative to calculate the rotational properties of NSs. The discovery of rapidly rotating pulsars like PSR J1748-2446ad has put strong upper bound on the maximum rotational frequency of NSs \cite{Hessels}. Moreover, in the slow rotation approximation ($P\leq 10 s$), theoretical constraints on the normalized moment of inertia have been obtained in terms of the tidal deformation \cite{Yagi} and compactness parameter \cite{Breu_Rez}. Such relations are called universal relations and are independent of EoS. Therefore it becomes much challenging to satisfy the aforesaid constraints on the various properties of neutron/compact star considering the formation of exotic matter like the hyperons, $\Delta$s and quarks. In the present work, I intend to investigate the possibility of hadron-quark phase transition and the gross properties of the resultant HS in the light of such constraints.
 
  To describe the hadronic phase the effective chiral model \cite{Sahu2004,TKJ} is adopted. In the present work, the hadronic phase consists of the baryon octet ($n,p,\Lambda,\Sigma^{-,0,+},\Xi^{-,0}$) and the delta quartet ($\Delta^{-,0,+,++}$) that interact via the $\sigma, \omega$ and $\rho$ mesons \cite{Sen4,Sen,Sen2,Sen3}. The model is well-tested and the parameters are determined on the basis of SNM properties \cite{Sahu2004,TKJ}. The detailed attributes of the model are discussed in \cite{Sahu2004,TKJ} while its salient features can be found in \cite{Sen4,Sen,Sen2,Sen3}. The highlights of the model and the parameter set adopted for the present work are discussed in the next sections \ref{Hadronic_model} and \ref{Model_param}, respectively. The set of hyperon and delta couplings are chosen same as in \cite{Sen4} and is also discussed briefly in the formalism section \ref{Couplings}.
  
  For the quark phase, the famous MIT bag model \cite{Chodos}, characterized by a bag constant $B$, has been employed. The inclusion of the repulsive quark interactions via the parameter $\alpha_4$ in the thermodynamic quark potential has been suggested by \cite{Glendenning,Weissenborn2011,Alford2005,Bombaci2017,Steiner,Khanmohamadi}. Hadron-quark phase transition in NS cores results in the formation of HSs. First order phase transition can be achieved using Gibbs construction (GC) or Maxwell construction (MC) depending on the surface tension ($\sigma_s$) at the hadron-quark phase boundary. However, as the value of $\sigma_s$ at the crossover boundary is still inconclusive, in the present work I proceed to achieve phase transition with GC assuming $\sigma_s$ to be small and compute the HS properties. It is to be remembered that the choice of $B$ and $\alpha_4$ play crucial roles in determining the phase transition properties like the threshold quark density, the density range of the mixed phase, the hybrid EoS as well as the structural properties of HSs. However, the limits to the values of $B$ and $\alpha_4$ are still indefinite. In the present work, I therefore choose moderate values of $B$ and $\alpha_4$, consistent with the prescriptions from GW170817 data, as mention in section \ref{Quark phase} of formalism section.

 This paper is organized as follows. The highlights of the hadronic model along with the model parameter and the hyperon and delta coupling scheme chosen for this work are discussed in section \ref{Hadronic_model}. A brief discussion on the attributes of the MIT bag model and the formalism for obtaining phase transition is added in section \ref{Quark phase}. A flavor of the mechanism to obtain the HS properties in both static and rotating conditions is added in section \ref{stat_props} and \ref{rot_props}, respectively. I present the results and their detailed analysis in the following section \ref{Results}. I finally put the conclusions of the present work in the last section \ref{conclusion}.

\section{Formalism}
\subsection{Hadronic matter with the baryon octet and delta quartet}
\label{Hadronic_model}

 The effective chiral model \cite{Sahu2004,TKJ,Sen5} has been adopted where the hyperons ($\Lambda,\Sigma^{-,0,+},\Xi^{-,0}$) and $\Delta$ baryons ($\Delta^{-,0,+,++}$) have been taken into account along with the nucleons, following the same formalism as \cite{Sen4}. It is a phenomenological model based on chiral symmetry with the scalar $\sigma$ and pseudo-scalar $\pi$ mesons being chiral partners. The spontaneous breaking of chiral symmetry at ground state leads to the dynamical generation of all the baryonic masses as well as that of the $\sigma$ and $\omega$ mesons \cite{Sen2,Sen4,Sen,Sen3,Sen5,Sahu2004,TKJ}. The model is also of RMF type and in such approximation, $<\pi> = 0$ and the pion mass becomes $m_{\pi} = 0$. Thus the pions do not contribute in this case. The isospin triplet $\rho$ mesons takes care of the isospin asymmetry in the system. An explicit mass term for the $\rho$ mesons is involved in the Lagrangian following\cite{Sahu2004,TKJ,Sen2,Sen4,Sen,Sen3,Sen5} though it is also possible to generate the mass of $\rho$ mesons dynamically like that of the $\sigma$ and $\omega$ mesons. Since the $\Delta^{-,0,+,++}$ baryons posses Breit-Wigner mass distribution, their mass is varied as (1232 $\pm$ 120) MeV \cite{Cai,Sun,Sen4} in the present work.

\subsubsection{The model parameter}
\label{Model_param}

\vspace*{-0.3cm}
 
 There are five model parameters $C_{\sigma N}$, $C_{\omega N}$, $C_{\rho N}$, $B$ and $C$. Of them $B$ and $C$ are the coefficients of higher order scalar field terms. The meson-nucleon couplings $g_{iN}$ are calculated in terms of $C_{iN}=g_i^2/m_i^2$, where $i=\sigma,\omega,\rho$. These parameters are obtained by reproducing the SNM properties \cite{TKJ}. The parameter set adopted for the present work is tabulated below in table \ref{table-1}. The same set has also been used in \cite{Sen,Sen2,Sen3,Sen4,Sen5}. 
 
\begin{table}[ht!]
\begin{center}
\caption{Model parameters chosen for the present work (adopted from \cite{TKJ}).}
\setlength{\tabcolsep}{15.0pt}
\begin{center}
\begin{tabular}{cccccccc}
\hline
\hline
\multicolumn{1}{c}{$C_{\sigma N}$}&
\multicolumn{1}{c}{$C_{\omega N}$} &
\multicolumn{1}{c}{$C_{\rho N}$} &
\multicolumn{1}{c}{$B/m^2$} &
\multicolumn{1}{c}{$C/m^4$} \\
\multicolumn{1}{c}{($\rm{fm^2}$)} &
\multicolumn{1}{c}{($\rm{fm^2}$)} &
\multicolumn{1}{c}{($\rm{fm^2}$)} &
\multicolumn{1}{c}{($\rm{fm^2}$)} &
\multicolumn{1}{c}{($\rm{fm^2}$)} \\
\hline
6.772  &1.995  & 5.285 &-4.274   &0.292 \\
\hline
\hline
\end{tabular}
\end{center}
\protect\label{table-1}
\end{center}
\end{table} 
 
 The values of SNM properties obtained with the above parameter set can be found in \cite{TKJ,Sen,Sen2,Sen3,Sen4,Sen5}. It is noteworthy that the nucleon effective mass $m_N^{\star}=0.85m_N$ for this model is quite high compared to other well-known RMF models and at high density unlike other RMF models, the effective mass increases after a certain value of density \cite{Sahu2004,TKJ,Sen2,Sen5}. This is because the effective mass of this model is dependent on both the scalar and vector fields and at high density the dominance of vector potential increases the nucleon effective mass. Moreover, the higher order terms of scalar field with coefficients $B$ and $C$ and the mass term of the vector field of the present model also become highly non-linear and dominant at high density. This leads to softening of EoS at high density \cite{Sahu2004,TKJ,Sen2,Sen5} and as seen from \cite{TKJ} the EoS for the adopted parameter set passes through the soft band of heavy-ion collision data. The EoS softens more when the formation of exotic baryons like hyperons and deltas are considered in NSM \cite{Sen,Sen2,Sen3,Sen4}. Thus the NS configurations, obtained in presence of such baryons, do not satisfy the maximum mass constraints for NSs \cite{Ant,Cromartie}. 
 
 Apart from nucleon effective mass, the other SNM properties like the binding energy per nucleon ($B/A=-16.3$ MeV) and the symmetry energy ($J=32$ MeV), the saturation density ($\rho_0=0.153~ \rm{fm}^{-3}$) match well with the estimates of \cite{Dutra2014,Stone}. The nuclear incompressibility ($K = 303$~ MeV) yielded by the chosen parameter set, is consistent with the results of \cite{Stone2} but it is larger than the estimate reported in \cite{Khan2012,Khan2013,Garg}. There are other parameters of the model shown in \cite{TKJ} that yield lower values of $K$ which are consistent with the range prescribed by \cite{Khan2012,Khan2013,Garg}. However, as discussed in \cite{Sen2,Sen5} such parameter sets cannot be adopted as they yield softer EoS \cite{TKJ} and consequently low mass NS configurations that do not satisfy the maximum mass constraint of NSs even with $\beta$ stable NSM. Therefore for the present work, I choose the parameter set (shown in table \ref{table-1}) following \cite{Sen,Sen2,Sen3,Sen4,Sen5} although it yields higher value of incompressibility compared to that prescribed in \cite{Khan2012,Khan2013,Garg}. Although the slope parameter ($L_0=87$ MeV) is a bit large compared to the findings of \cite{Tsang}, it is quite consistent with the range specified by \cite{Dutra2014}. Moreover, recent co-relation between the symmetry energy and tidal deformability and radius of a 1.4 $M_{\odot}$ NS shows that $L_0$ can be as high as $\sim 80$ MeV \cite{Fattoyev,Zhu2018}. 
 
 This model and the adopted parameter set is thus well-tested to describe nuclear matter at finite temperature \cite{Sen5} and hadron-quark phase transition with $\Delta$ baryons \cite{Sen} and hyperon matter in HSs both in static and rotating cases \cite{Sen2}.

\subsubsection{Hyperon and delta couplings}
\label{Couplings}
\vspace*{-0.3cm}

 Similar to \cite{Weissenborn2012,TKJ2,TKJ3,Gupta,Sen2,Sen3,Sen4}, the hyperon-meson couplings $x_{iH}=g_{iH}/g_{iN}$ (where, $i=\sigma,\omega,\rho$ and $H=\Lambda,\Sigma,\Xi$) are calculated following the constraint ($x_{\sigma H} \leq 0.72$ \cite{Glendenning,Glen2,Rufa}) from hyper-nuclear studies on the scalar couplings $x_{\sigma H}$ while the corresponding vector couplings $x_{\omega H}$ can be obtained in terms of the potential depths of the individual hyperon species ($(B/A)_H |_{\rho_0}$ = -28 MeV for $\Lambda$, +30 MeV for $\Sigma$ and -18 MeV for $\Xi$ \cite{Schaffner-Bielich,Sulaksono,Ishizuka,Sen2,Sen3,Sen4}). Among them $(B/A)_{\Lambda} |_{\rho_0}$ is known by extrapolating the $\Lambda$ binding energy of finite hypernuclei in the limit of infinite matter. In the present work, $x_{\sigma H}=0.7$ following \cite{Sen4}. Following \cite{Sen2,Sen3,Sen4}, $x_{\rho H}$ is chosen same as $x_{\omega H}$ due to the similar mass values of $\rho$ and $\omega$ mesons and also because both are responsible for the generation of short range repulsive forces.
 
 However, the potential depth of the $\Delta$s is still experimentally poorly determined. However, \cite{Drago2014,Riek,Kolomeitsev,Maslov} have suggested different possible range in this regard. Therefore the $\Delta$-meson couplings $x_{i\Delta}=g_{i\Delta}/g_{iN}$ (where, $i=\sigma,\omega,\rho$) are inconclusive at present. However, there are certain suggestions based on theoretical perspectives \cite{Boguta1982,Lavagno,Kosov} and QCD calculations \cite{Jin}. A detailed discussion on the uncertainties pertaining to the $\Delta$ couplings can be found in \cite{Sen}. In the present work the $\Delta$-meson couplings are chosen following the constraint prescribed by \cite{Boguta1982,Lavagno}. There is no definite suggestion regarding the choice of $x_{\rho \Delta}$. Ref. \cite{Sen} shows that considering the present model, $\Delta$s are not formed in NSM if $x_{\rho \Delta} \geq 1$. Thus similar to \cite{Sen4,Sen} the delta coupling set is chosen as ($x_{\sigma \Delta},x_{\omega \Delta},x_{\rho \Delta}$)=(1.35,1.0,1.0). It can be seen from \cite{Sen} that this delta coupling set yields delta potential (-110 MeV) which is consistent to the range suggested by \cite{Kolomeitsev,Maslov}. Also with this set the second minima of the energy per baryon (8 MeV at 2.5$\rho_0$) lie well above the saturation energy of normal nuclear matter (-16.3 MeV at $\rho_0$) \cite{Sen,Sen4}. This respects the criteria for choosing $\Delta$ couplings prescribed by \cite{Boguta1982,Lavagno,Kosov}.

 With the above mentioned couplings, the hadronic EoS is computed for three values of the delta baryon mass $m_{\Delta}=1112,1232 ~\& 1352$ MeV.
 
\subsection{Pure quark phase \& hadron-quark phase transition}
\label{Quark phase}

 The well-known MIT bag model \cite{Chodos} is taken into account to describe the pure quark phase consisting of the unpaired u, d and s quarks along with the electrons. The bag constant $B$ and repulsive parameter $\alpha_4$ determine the strength of strong repulsive interaction between the quarks \cite{Alford2005,Fraga,Schramm,Bombaci2017,Weissenborn2011,Sen}. The thermodynamic quark potential of such a system is given by \cite{Weissenborn2011,Glendenning,Sen} from which one can obtain the EoS for the pure quark phase. The u and d quark masses are much smaller compared to that of the s quark ($m_s$=100~MeV) \cite{Nakazato,PDG}. As mentioned in the introduction section \ref{intro}, the values of $B$ and $\alpha_4$ are still uncertain. It is also known that the perturbative effects on HS properties can also be realized by changing the value of $B$\cite{Steiner,Prakash,Yazdizadeh,Burgio,Miyatsu2015,Liu}. The higher these values the stiffer is the EoS giving more massive HS configurations \cite{Bag,Yudin,Logoteta2}. Literature suggests $B^{1/4}\sim ((100-300)$ MeV)$^4$ \cite{Steiner,Buballa,Novikov,Baym} while lattice calculations predicts $B^{1/4}\sim 210$ MeV/fm$^3$ \cite{Benhar}. Recently, consistent with GW170817 observation and measurement of $\Lambda_{1.4}$ \& $R_{1.4}$, \cite{EnPingZhou} suggests that $B^{1/4} = (134.1-141.4)$ MeV and $\alpha_4 = (0.56-0.91)$ for a low-spin prior while for the high-spin priors $B^{1/4} = (126.1-141.4)$ MeV and $\alpha_4 = (0.45-0.91)$ considering pure quark stars. Ref. \cite{Nandi} suggested similar maximum values of $B^{1/4}$ and $\alpha_4$ for HSs while \cite{Rather} also suggests $B^{1/4}=(130-160)$ MeV. Consistent with recent prescriptions from GW170817 data analysis \cite{Nandi,Rather}, the hybrid EoS is obtained in the present work by choosing moderate values as $B^{1/4}=150$ MeV and $\alpha_4=0.5$.

 As mentioned in the introduction section \ref{intro}, GC or MC can be employed to achieve phase transitions in HS cores, depending on the value of surface tension ($\sigma_s$) at the hadron-quark phase boundary. According to \cite{Maruyama,Maruyama2,Endo,Sotani,Shahrbaf,Xia2019} if $\sigma_s \lesssim 70$ MeV fm$^{-3}$, GC is favored with the formation of an intermediate stable mixed phase where both hadronic and quark matter coexist \cite{Glendenning,Glenq,Logoteta2,Orsaria2013,Rotondo,Bhattacharya,Sen,Sen2}. However, if the value of $\sigma_s$ is higher, the mixed phase becomes unstable and MC is then adopted to obtain phase transition. Therefore, unlike MC, a mixed phase region is expected in case of GC rather than a density jump as in case of MC \cite{Schramm,Lenzi,Bhattacharya,Logoteta2013,Sen,Sen2,Gomes2019,Han,Ferreira2020}. This is because in case of GC, both the neutron and electron chemical potentials are continuous along with pressure but MC predicts the continuous variation of pressure and neutron chemical potential with jump in electron chemical potential at the transition density. Moreover, GC is ruled by global charge neutrality condition that states the overall HS matter (HSM) must be charge neutral unlike MC which states that the pure hadronic and quark phases must be individually charge neutral \cite{Bhattacharya,Sen,Sen2,Gomes2019,Han}. In case of GC with low values of surface tension, the presence of quark matter in HSs enables the hadronic regions of the mixed phase to become more isospin symmetric than in the pure phase by transferring electric charge to the quark phase \cite{Sen2}. In the present work, phase transition is obtained by assuming the surface tension at hadron-quark boundary to be small and hence using GC.

 Once the hybrid EoS is obtained, the speed of sound ($C_s$) can also be calculated as the first order derivative of pressure with respect to energy density \cite{Han,Whittenbury2015}. It is known that the speed of sound plays important role in the context of phase transition. It is suggested that under such circumstances, the value $C_s$ can vary drastically and may surpass the conformal limit of $C_s^2=1/3$ and be close to the causality limit of $C_s^2<1$ \cite{cs3,Tews}. Therefore, it will be interesting to study the effects of phase transition on the speed of sound in the present work.

\subsection{Neutron Star Structure \& Properties}

 As the structural properties of NSs/HSs depend solely on the EoS, the former can be obtained with the computed hybrid EoS.  
 
\subsubsection{Static properties}
\label{stat_props}
\vspace*{-0.3cm}

 For the hybrid EoS, the structural properties of HSs are computed in static conditions by integrating the Tolman-Oppenheimer-Volkoff (TOV) equations \cite{tov,tov2} that depict the hydrostatic equilibrium between gravity and the internal pressure of the star. Solving these equations, the static properties like central energy density ($\varepsilon_c$), gravitational ($M$) and baryonic masses ($M_B$), radius ($R$) and the surface redshift ($Z_s$) of the HS are calculated for the obtained hybrid EoS.


\subsubsection{Rotational properties}
\label{rot_props}
 \vspace*{-0.3cm}

 The rotational properties like rotational mass ($M$) and radius ($R$), maximum rotational frequency ($\nu_k$) and the moment of inertia ($I$) are calculated using the RNS code \cite{RNS}.

\section{Result and Discussions}
\label{Results}

\subsection{Neutron Star properties with pure hadronic matter}

 The hadronic EoS is constructed by varying the $\Delta$ baryon mass $m_{\Delta}$ in presence of hyperons and following the coupling scheme discussed in the formalism section (\ref{Couplings}) for the hyperons and $\Delta$s. The EoS and the formation of the baryons in NSM can be found in \cite{Sen4} for the same set of hyperons and $\Delta$ couplings with the same hadronic model. EoS softens quite a lot due to considerable formation of hyperons and $\Delta$s and the softening is maximum when $m_{\Delta}$ is minimum and therefore most favored in NSM. The $\Delta$ baryon mass indeed plays an important role in the population fraction of various particles and the EoS. Interestingly, for minimum value of $\Delta$ baryon mass (1112 MeV), the formation of hyperons is completely suppressed by the $\Delta$s of all charge whereas when the maximum delta mass (1352 MeV) is considered, it is the $\Delta$s that get completely suppressed by the hyperons. The intermediate mass value of $\Delta$s (1232 MeV) yields formation of both hyperons and $\Delta$s with comparatively fast deleptonization \cite{Sen4}.

\subsection{Hadron-quark phase transition}

\begin{figure}[!ht]
\centering
\includegraphics[scale=1.3]{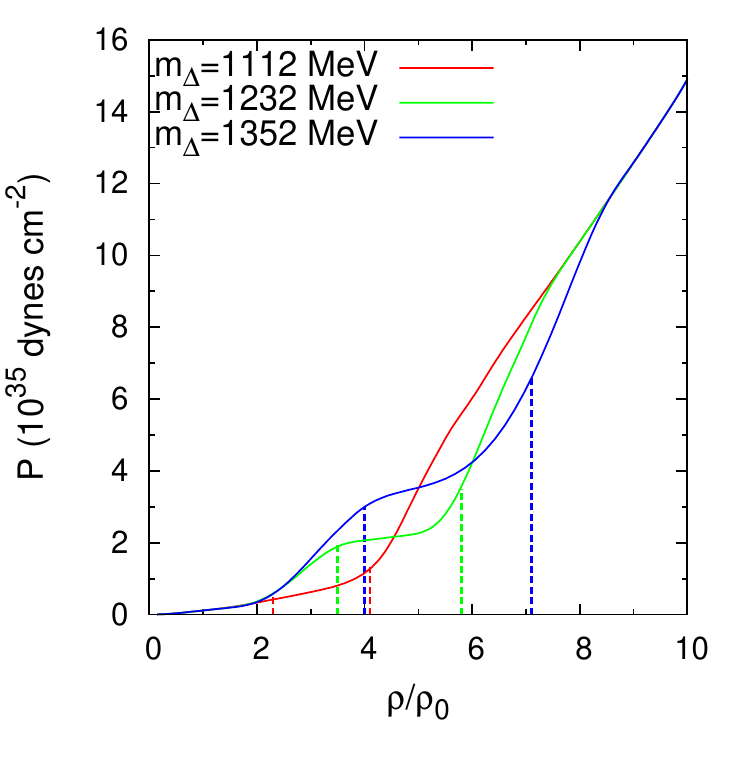}
\caption{\label{eosGC} Equation of State ($\varepsilon ~vs.~ P$) of hybrid star matter for different masses of $\Delta$ baryons. The hadron-quark mixed phase regions are also marked by vertical dashed lines.}
\end{figure}

 Using the chosen values of $B$ and $\alpha_4$, discussed in the formalism section (\ref{Quark phase}), the hybrid EoS is constructed with GC and presented in figure \ref{eosGC}. 
 
 Figure \ref{eosGC} shows smooth phase transition with GC characterized by stable and distinct mixed phases for all the hybrid EoS. For $m_{\Delta}=1112$ MeV, the mixed phase ranges as $(2.3-4.1)\rho_0$ while it starts from $3.5 \rho_0$ and ends at $5.8\rho_0$ for $m_{\Delta}=1232$ MeV. For $m_{\Delta}=1352$ MeV, the mixed phase exists from $4.0\rho_0$ to $7.1\rho_0$. The stiffer the hadronic EoS, the delayed is the transition and more is the stretch of the mixed phase region.
 
\begin{figure}[!ht]
\centering
\includegraphics[scale=0.8]{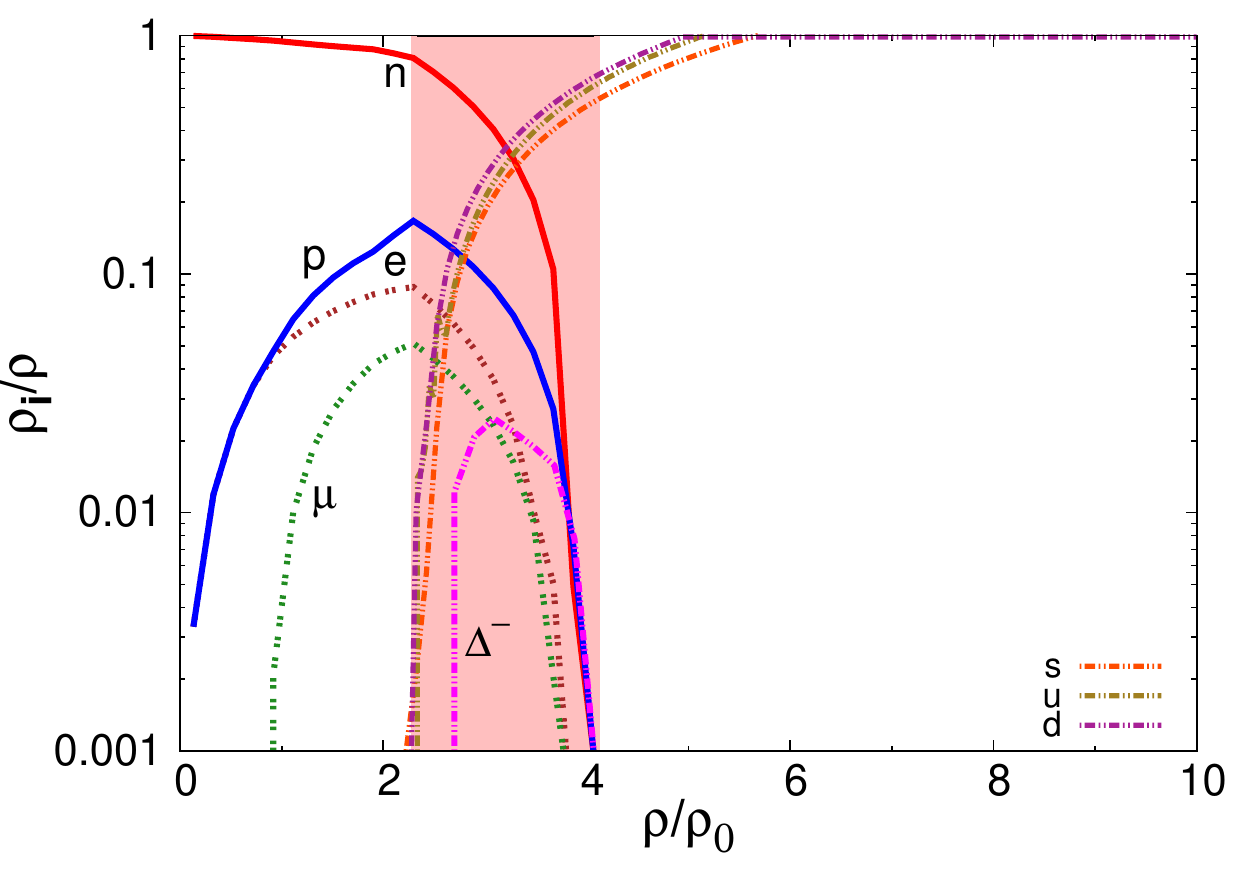}
\caption{\label{pfHq1} Relative particle fraction of different baryons and quarks in hybrid star matter for $m_\Delta$ = 1112 MeV. The shaded region indicates the mixed phase.}
\end{figure}

\begin{figure}[!ht]
\centering
\includegraphics[scale=0.8]{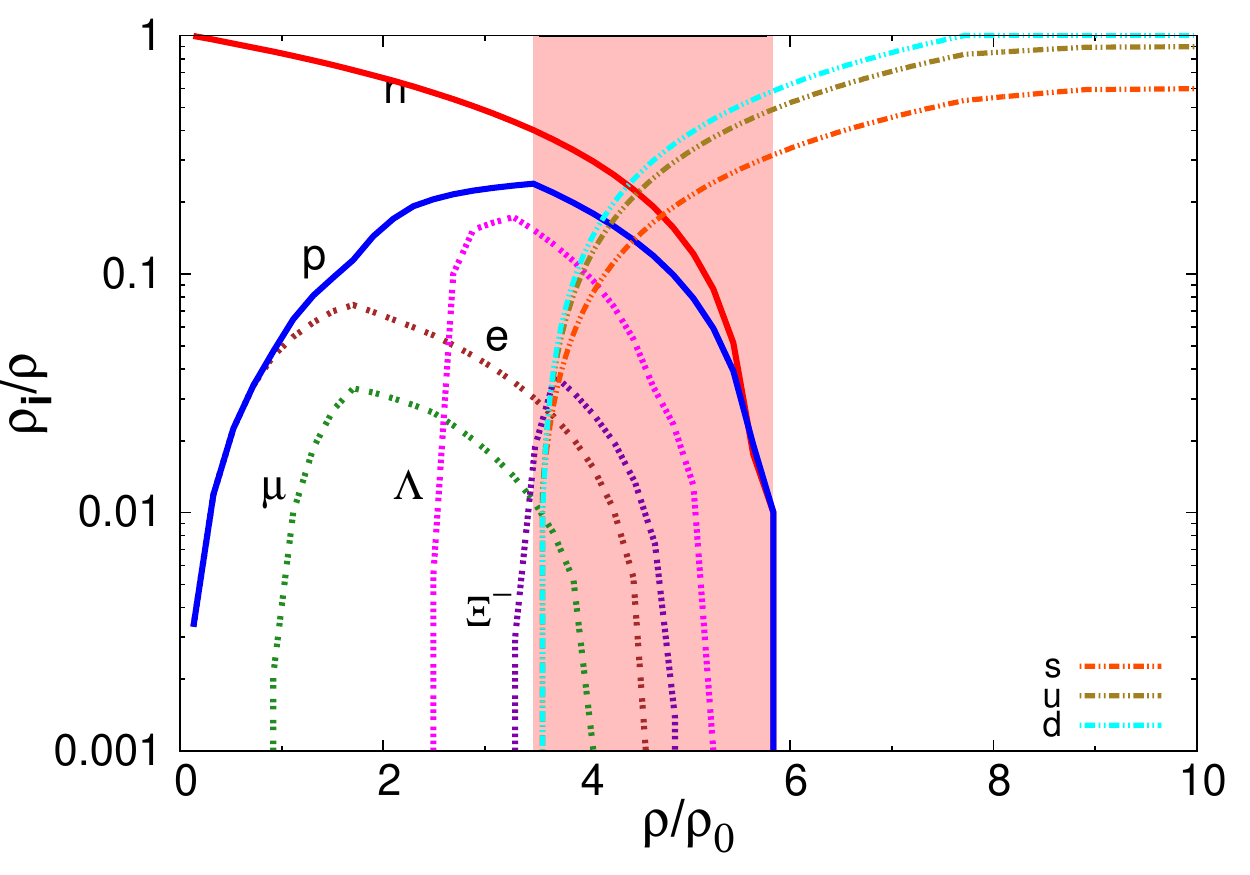}
\caption{\label{pfHq2} Same as \ref{pfHq1} but for $m_\Delta$ = 1232 MeV.}
\end{figure}

\begin{figure}[!ht]
\centering
\includegraphics[scale=0.85]{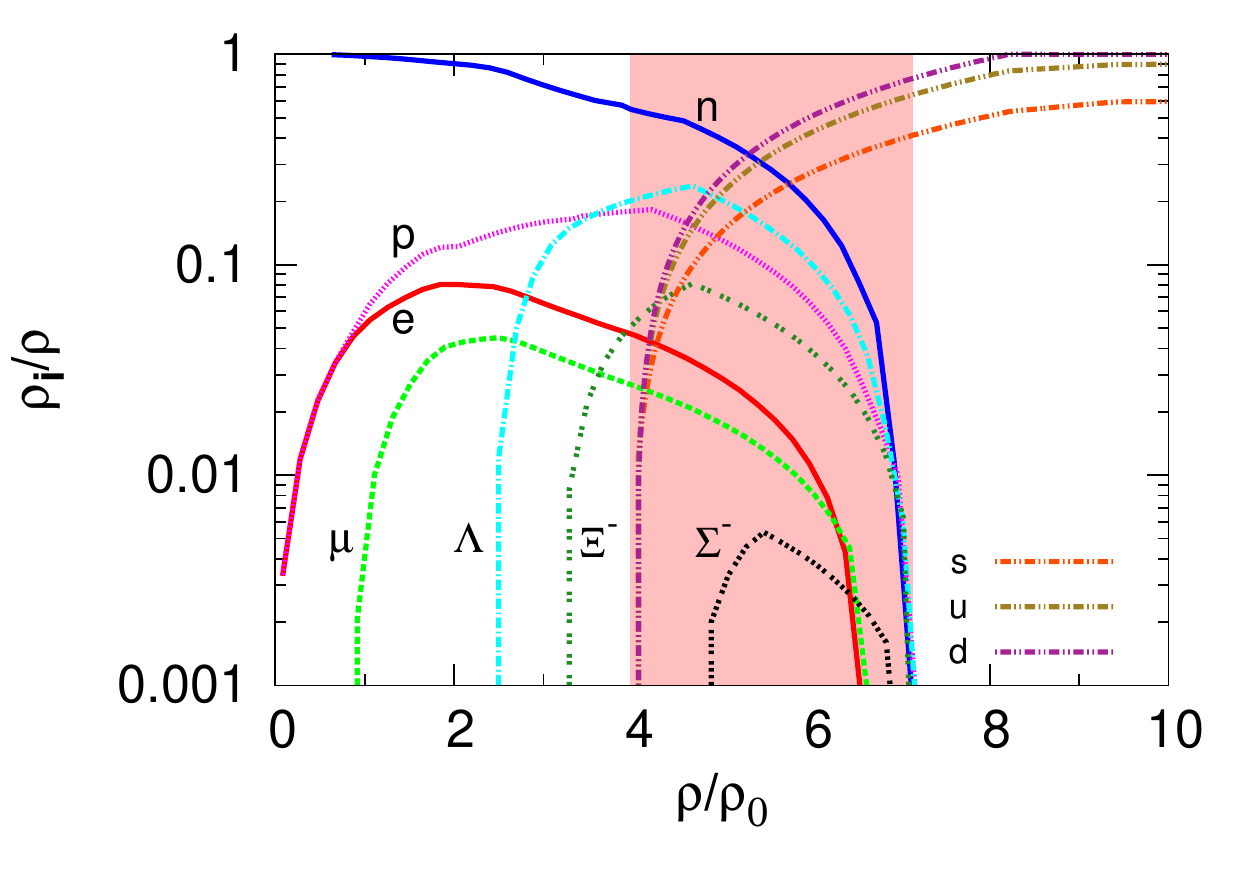}
\caption{\label{pfHq3} Same as \ref{pfHq1} but for $m_\Delta$ = 1352 MeV.}
\end{figure}

 The relative abundance of hadrons and quarks in the HSM is studied and shown in figures \ref{pfHq1},\ref{pfHq2} and \ref{pfHq3} for $m_{\Delta}=$1112, 1232 and 1352 MeV, respectively. 

 The early formation of quarks in case of $m_\Delta$ = 1112 MeV suppresses the formation of hadrons to a large extent. Only a feeble fraction of $\Delta^-$ is formed along with the nucleons and leptons (figure \ref{pfHq1}). For the other two values of $m_\Delta$, no $\Delta$ baryons appear in the HSM. The formation of hyperons are also not much favorable in HSM except for the $\Lambda$ and $\Xi^-$ which populate HSM in considerable amount (figures \ref{pfHq2} and \ref{pfHq3}). It is noteworthy that $\Xi^-$, though heavier than the $\Sigma$s, appears earlier because the latter having positive potential depth (+30 MeV) suffer more repulsion at high density \cite{Sen2,Sen3,Sen4,Glen2001}. Consistent with \cite{Bhattacharya,Miyatsu2015}, it is seen that the population of down quarks are predominant in HSM, followed by that of the up quarks and finally the strange quarks. Although the d and s quark are treated in the same way with equal chemical potential and both have the same charge, the s quarks being massive compared to the other two, it has comparatively less concentration in HSM.
 
\begin{figure}[!ht]
\centering
\includegraphics[scale=1.4]{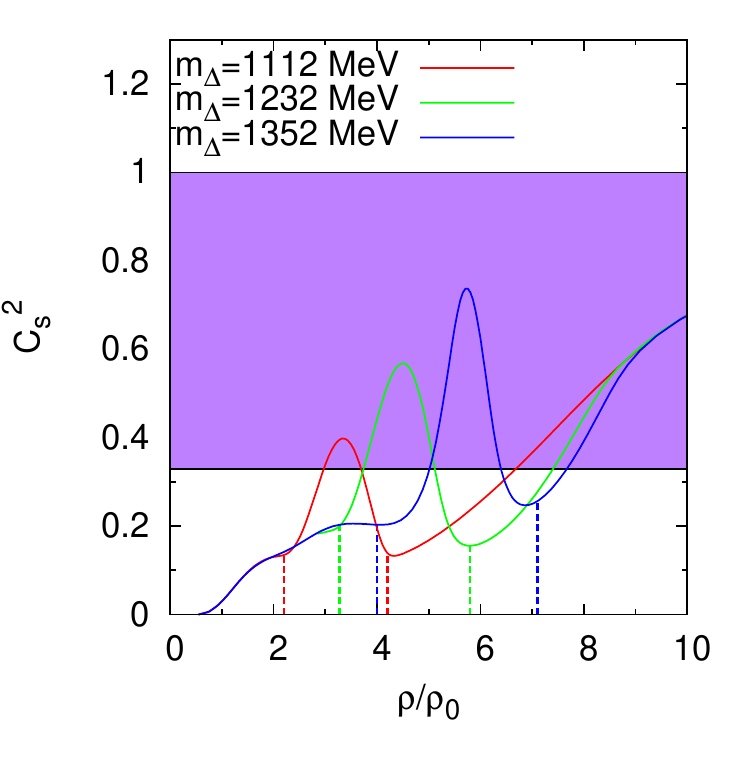}
\caption{\label{Cs} Speed of sound in hybrid star matter for different masses of $\Delta$ baryons. The mixed phase regions are also marked with vertical dashed lines. The shaded region shows the bound on speed of sound from GW170817 analysis \cite{Kanakis-Pegios}.}
\end{figure}

 The speed of sound and its behavior is studied in HSM and depicted in figure \ref{Cs}.

 The density dependence of speed of sound shows that it increases monotonically for the pure hadronic phase. Unlike the case of MC where the value of $C_s^2$ drops to zero in the phase transition region \cite{Castillo,Blaschke}, it is seen that in the mixed phase region with GC, the initial increase of $C_s^2$ is rapid and drastic. However, it again decreases sharply as the mixed phase region ends to initiate pure quark phase, thereby $C_s^2$ showing a peak value within the mixed phase region. The result is consistent with that of \cite{MiyatsuPoS,cs3,Whittenbury2015}. The speed of sound shows monotonic increase in the pure quark phase. As expected the speed of sound is more in case of quark matter. For $m_\Delta$ = 1112 MeV, the peak (0.40) is observed at $3.3 \rho_0$ while for $m_\Delta$ = 1232 MeV and $m_\Delta$ = 1352 MeV, the maximum values of $C_s^2$ are 0.57 ($4.5 \rho_0$) and 0.80 ($5.7 \rho_0$). It is noteworthy that the peaks of $C_s^2$ with all the hybrid EoS are obtained within the bound (lower bound = $1/3$ and upper bound = $1$ in $c=1$ units) specified by \cite{Kanakis-Pegios,NaZhang,cs3,Reed,Marczenko} consistent with GW170817 observations.

\subsection{Static hybrid stars}

\begin{figure}[!ht]
\centering
\includegraphics[scale=1.2]{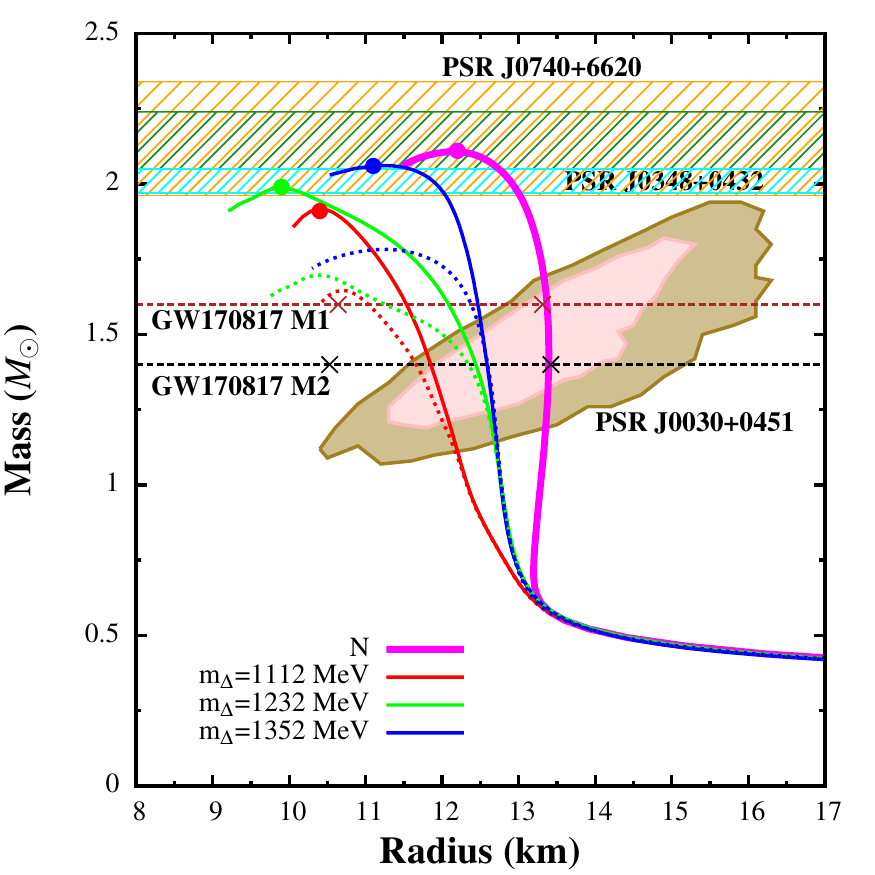}
\caption{\label{mr_all} Mass-Radius relationship for static stars with $\beta$ stable matter (denoted by `N' (thick solid magenta curve)), in presence of hyperons and deltas (dotted curves) and hybrid star matter (thin solid curves) for different masses of $\Delta$ baryons. Observational limits imposed from high mass pulsars like PSR J0348+0432 ($M = 2.01 \pm 0.04~ M_{\odot}$) \cite{Ant} (cyan shaded region) and PSR J0740+6620 ($2.14^{+0.10}_{-0.09}~ M_{\odot}$ (68.3\% - dark green shaded region) and $2.14^{+0.20}_{-0.18}~ M_{\odot}$ (95.4\% - orange shaded region)) \cite{Cromartie} are also indicated. The horizontal black and brown dashed horizontal lines indicate the canonical mass $M = 1.4~M_{\odot}$ (GW170817 M1) and mass $M = 1.6~M_{\odot}$ (GW170817 M2), respectively. The limits on $R_{1.4}$ \cite{Abbott,Fattoyev} and $R_{1.6}$ \cite{Bauswein} prescribed from GW170817 are indicated by crossmarks. The constraints on M-R plane from NICER experiment for PSR J0030+0451 \cite{Miller} are also compared (95\% - outer grey shaded region and 68\% - inner pink shaded region).}
\end{figure}

 The static properties of HSs like central energy density ($\varepsilon_c$), gravitational ($M$) and baryonic masses ($M_B$), radius ($R$) and the surface redshift ($Z_s$) are obtained solving the TOV equations. In figure \ref{mr_all}, the variation of gravitational mass with radius is shown for different values of $m_{\Delta}$.

 The maximum gravitational mass for $\beta$ stable star (N) is found to be 2.10 $M_{\odot}$ with corresponding radius 12.2 km. The hadronic matter solutions (in presence of hyperons and $\Delta$s) \cite{Sen4} are also compared to infer that with phase transition, there is considerable increase in the gravitational mass $\approx(15-17)$\%. This result is consistent with that of \cite{Ozel2010,Weissenborn2011,Klahn,Bonanno,Lastowiecki,Drago2016,Drago2016(2),Bombaci2016, Bombaci2017}. Like few other relativistic models \cite{Bhowmick,Ishizuka}, none of the solutions obtained with hadronic matter EoS in presence of hyperons and $\Delta$s (dotted lines) satisfy the maximum mass constraint obtained from the most massive pulsars like PSR J0348+0432 \cite{Ant} and PSR J0740+6620 \cite{Cromartie}. On the other hand, with hybrid EoS, for both $m_{\Delta}=1232$ MeV (1.98 $M_{\odot}$) and $m_{\Delta}=1352$ MeV (2.06 $M_{\odot}$), the maximum mass constraints from both PSR J0348+0432 and PSR J0740+6620 are satisfied. However, for $m_{\Delta}=1112$ MeV the obtained value of maximum gravitational mass is little low (1.91 $M_{\odot}$). All the $M-R$ solutions shown in figure \ref{mr_all} are in excellent agreement with the results from NICER experiment for PSR J0030+0451 \cite{Miller}. Also the radii values $R_{1.4}$ and $R_{1.6}$ are in good agreement with that suggested from the analysis of GW170817 data \cite{Abbott,Fattoyev,Bauswein}.

\begin{figure}[!ht]
\centering
\includegraphics[scale=1.2]{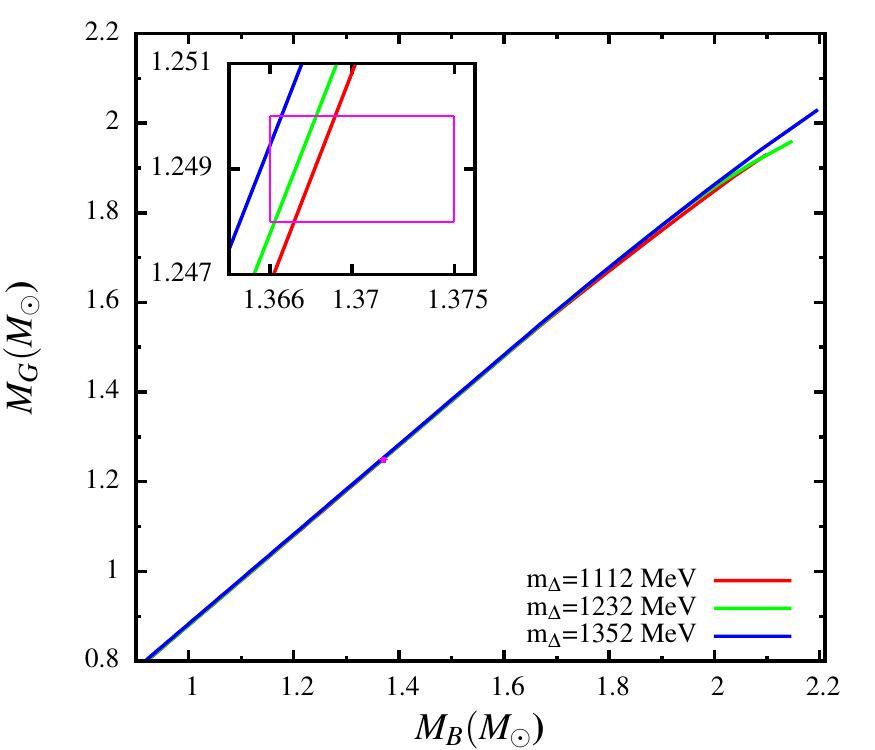}
\caption{\label{mgmb} Baryonic mass vs gravitational mass of hybrid stars for different masses of $\Delta$ baryons. The magenta box represent the constraint of \cite{Podsiadlowski} on baryonic mass ($M_B = (1.366 - 1.375)M_{\odot}$) for Pulsar B of binary system PSR J0737-3039 with gravitational mass ($M_G = (1.249 \pm 0.001)M_{\odot}$) \cite{Burgay}}.
\end{figure}

 Next the variation of gravitational mass is depicted with respect to the baryonic mass in figure \ref{mgmb}.

 The inset of figure \ref{mgmb} shows that with all the hybrid EoS, the constraint on baryonic mass from PSR J0737-3039B ($M_B = (1.366 - 1.375) M_{\odot}$) \cite{Podsiadlowski} with corresponding maximum gravitational mass $M_G = (1.249 \pm 0.001) M_{\odot}$ \cite{Burgay} has been satisfied. Consistent with results of \cite{Kolomeitsev,Maslov,Drago2016}, it is seen that this constraint is better satisfied as more $\Delta$ baryons are formed when they are considered to be less massive.
 
\begin{figure}[!ht]
\centering
\includegraphics[scale=1.2]{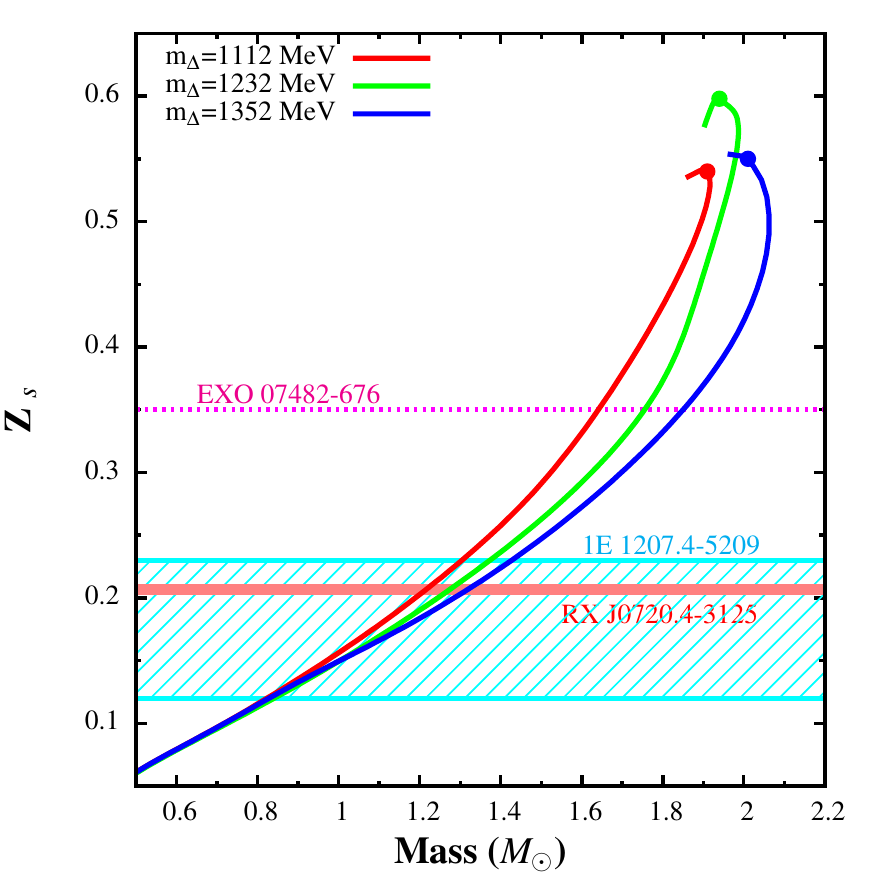}
\caption{\label{mZ} Surface gravitational redshift vs mass of hybrid stars for different masses of $\Delta$ baryons. Observational limits imposed from EXO 07482-676 ($Z_S = 0.35$) \cite{Cottam2002}, 1E 1207.4-5209 ($Z_S = (0.12 - 0.23)$) \cite{Sanwal} and RX J0720.4-3125 ($Z_S = 0.205_{-0.003}^{+0.006}$ \cite{Hambaryan} are also indicated.}
\end{figure}

 The surface redshift is calculated in terms of $M$ and $R$. Figure \ref{mZ} shows the variation of $Z_s$ with $M$.

 Figure \ref{mZ} depicts that the redshift is maximum (0.57) for $m_{\Delta}=1232$ MeV although the mass is maximum for $m_{\Delta}=1352$ MeV. This is because $Z_s$ depends both on the mass and radius. The hybrid EoS for $m_{\Delta}=1232$ yields the minimum radius (9.91 km) corresponding maximum mass 1.98 $M_{\odot}$ compared to that for $m_{\Delta}=1352$ (11.15 km; 2.06 $M_{\odot}$). For $m_{\Delta}=1112$ MeV and $m_{\Delta}=1232$ MeV, the maximum values of $Z_s$ are 0.54 and 0.55, respectively. The values of maximum redshift obtained with all the hybrid EoS satisfy the observational bounds obtained from EXO 07482-676 \cite{Cottam2002}, 1E 1207.4-5209 \cite{Sanwal} and RX J0720.4-3125 \cite{Hambaryan}.
\\
 The various static properties of HSs obtained from the present analysis are tabulated in table \ref{Stat.Prop}.

\begin{table}[ht!]
\caption{Static properties of stars with $\beta$ stable matter (N), hadronic matter in presence of hyperons and deltas (H) and hybrid star matter (HSM) for different masses of $\Delta$ baryons.}
\begin{center}
\begin{tabular}{cccccccccc}
\hline
\hline
\multicolumn{1}{c}{$m_{\Delta}$}&
\multicolumn{1}{c}{}&
\multicolumn{1}{c}{$M$}&
\multicolumn{1}{c}{$M_{B}$} &
\multicolumn{1}{c}{$R$} &
\multicolumn{1}{c}{$R_{1.4}$} &
\multicolumn{1}{c}{$R_{1.6}$} & \\
\multicolumn{1}{c}{(MeV)} &
\multicolumn{1}{c}{} &
\multicolumn{1}{c}{($M_{\odot}$)} &
\multicolumn{1}{c}{($M_{\odot}$)} &
\multicolumn{1}{c}{($km$)} &
\multicolumn{1}{c}{($km$)} &
\multicolumn{1}{c}{($km$)} & \\
\hline
\hline
-    &N    &2.10 &2.41 &12.2 &13.4 &13.3 \\
\hline
1112 &H    &1.65 &1.77 &10.7 &11.6 &11.0 \\
     &HSM  &1.91 &2.10 &10.4 &11.8 &11.5 \\
\hline     
1232 &H    &1.69 &1.82 &10.4 &12.3 &11.2 \\
     &HSM  &1.98 &2.15 &9.9  &12.4 &12.1 \\
\hline     
1352 &H    &1.76 &1.96 &11.2 &12.5 &12.4 \\
     &HSM  &2.06 &2.19 &11.1 &12.6 &12.5 \\
\hline
\hline
\end{tabular}
\end{center}
\protect\label{Stat.Prop}
\end{table}

\subsection{Rotating hybrid stars}

 The rotational properties of HSs like the rotational mass, radius, central energy density, rotational frequency and the moment of inertia are calculated using the RNS code \cite{RNS}.
 
\begin{figure}[!ht]
\centering
\includegraphics[scale=0.72]{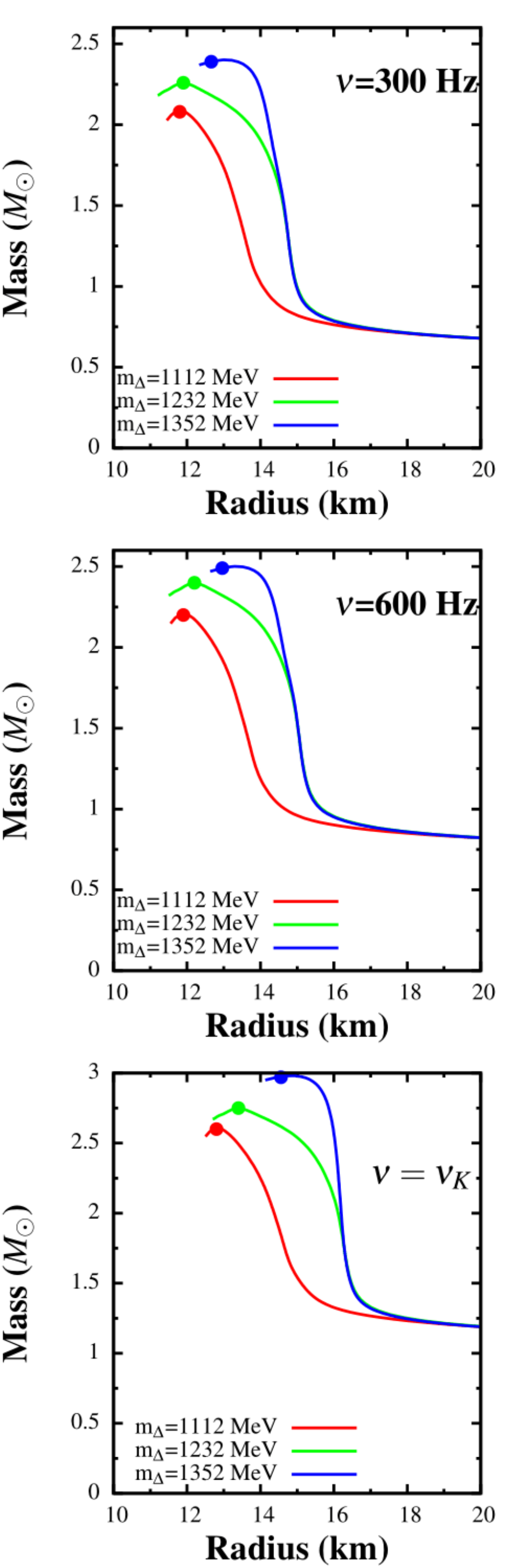}
\caption{\label{RotAll} Mass-Radius relationship of hybrid stars for different masses of $\Delta$ baryons rotating with different rotational frequencies.}
\end{figure}

 The rotational mass and radius of the HSs for different values of $m_{\Delta}$ are calculated for rotational frequencies $\nu=300, 600$ Hz and the Kepler frequency $\nu_K$. The results are presented in figure \ref{RotAll}.

 Both gravitational mass and radius increase with rotational frequency (angular
velocity). This is because of the centrifugal force that increases with increasing frequency and affects greatly the rotational mass and radius of NSs. Thus for any given value of $m_{\Delta}$, both $M$ and $R$ are maximum at Kepler frequency which is the maximum frequency of a stable rotating NS.

\begin{figure}[!ht]
\centering
\includegraphics[scale=1.2]{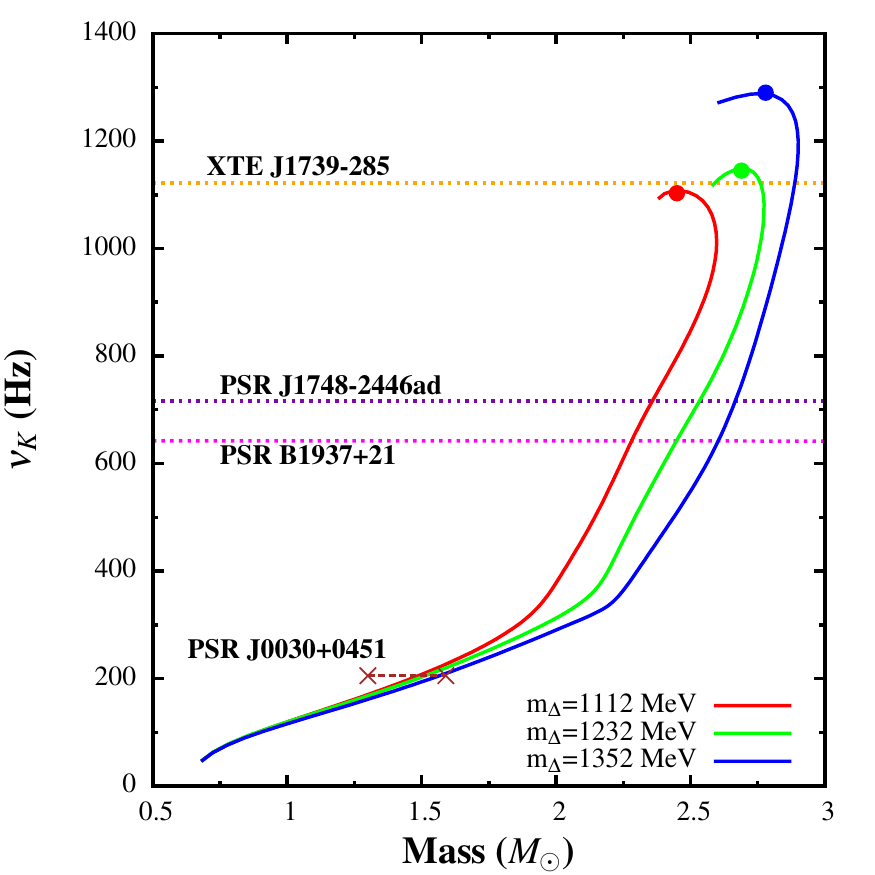}
\caption{\label{mn} Rotational frequency versus gravitational mass for hybrid star for different masses of $\Delta$ baryons rotating at Kepler velocity. The frequencies from fast rotating pulsars such as PSR B1937+21 ($\nu = 633$ Hz) \cite{Backer} and PSR J1748-2446ad ($\nu = 716$ Hz) \cite{Hessels} and XTE J1739-285 ($\nu=1122$ Hz) \cite{Kaaret} are also indicated. Range of gravitational mass ($M=1.44^{+0.15}_{-0.14}~ M_{\odot}$) for PSR J0030+0451 with $\nu=205.53$ Hz \cite{Miller} is shown with the brown dotted line and crossmarks}.
\end{figure}

 Figure \ref{mn} shows the variation of rotational frequency of HSs with respect to gravitational mass at Keplerian velocity. 

 As expected, massive NSs can experience fast rotation. Therefore the rotational frequency is maximum (1297 Hz) in case of the most massive HS configuration obtained with $m_{\Delta}=1352$ MeV. For $m_{\Delta}=1112$ MeV and $m_{\Delta}=1232$ MeV, the values of maximum rotational frequency are 1103.5 and 1144.4 Hz, respectively. The estimates of rotational frequency for all the HS configuration satisfy the observational constraints from PSR B1937+21 \cite{Backer} and PSR J1748-2446ad \cite{Hessels}. For both $m_{\Delta}=1232$ MeV and $m_{\Delta}=1352$ MeV, the bound from XTE J1739-285 \cite{Kaaret} is satisfied. Also for slow rotation ($\nu=205.53$ Hz), the obtained values of gravitational mass with the three HS configurations, lie within the range specified from PSR J0030+0451 \cite{Miller}.
 
\begin{figure}[!ht]
\centering
\includegraphics[scale=1.2]{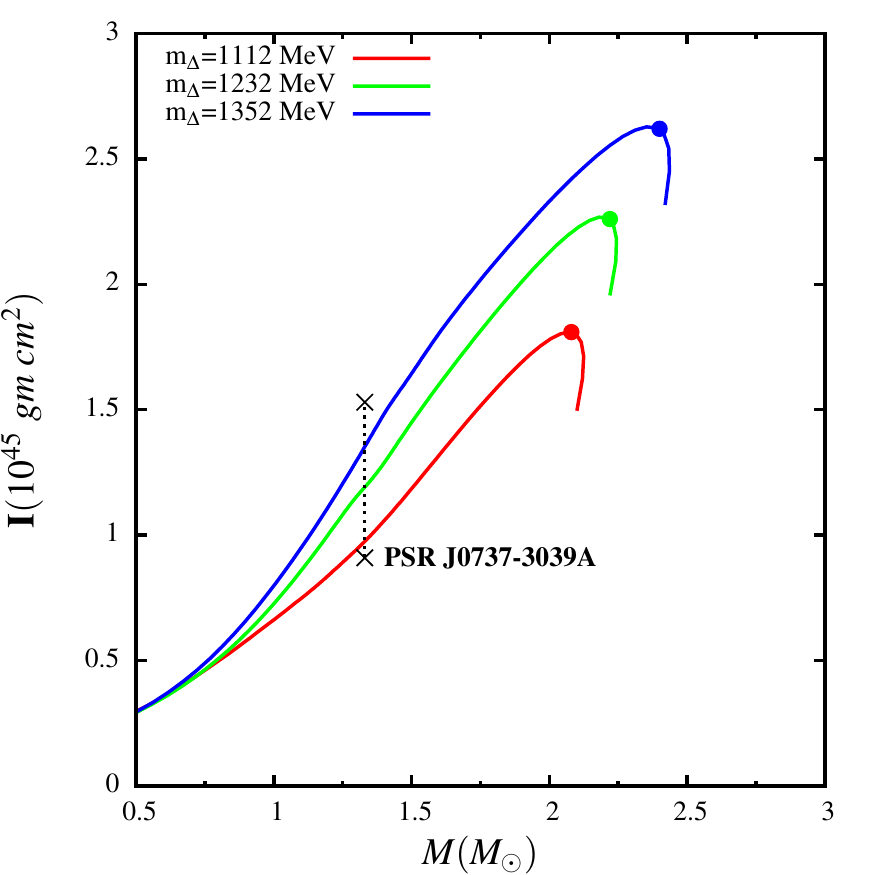}
\caption{\label{mI300} Moment of inertia ($I$) as a function of gravitational mass ($M$) of hybrid stars for different masses of $\Delta$ baryons rotating with frequency $\nu=300$ Hz. Constraint from PSR J0737-3039A ($I=1.15^{+0.38}_{-0.24}$ $\times 10^{45}$ g cm$^{2}$ for $M=1.338 M_{\odot}$) \cite{Bharat} is also shown}.
\end{figure}

\begin{figure}[!ht]
\centering
\includegraphics[scale=1.2]{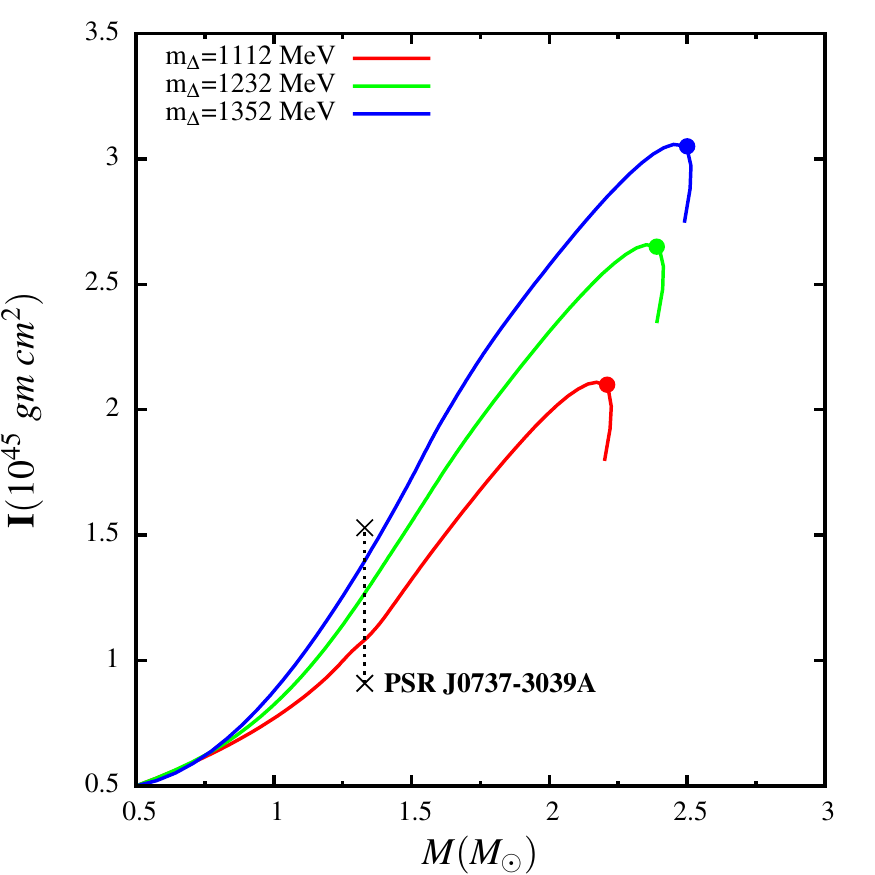}
\caption{\label{mI600} Same as figure \ref{mI300} but for $\nu=600$ Hz.}.
\end{figure}

 Next the moment of inertia profile for slow rotation is studied with the three hybrid EoS. The change of $I$ with respect to $M$ is shown in figures \ref{mI300} and \ref{mI600} for rotational frequencies $\nu=300$ and $\nu=600$ Hz, respectively.

 As expected, the moment of inertia is larger for massive stars as they can sustain faster rotation. Therefore moment of inertia also increases with rotational speed for a given hybrid EoS. It can be seen from figures \ref{mI300} and \ref{mI600} that all the hybrid EoS satisfy the slow rotational constraint from PSR J0737-3039A \cite{Bharat} for both $\nu=300$ and $\nu=600$ Hz.
 
\begin{figure}[!ht]
\centering
\includegraphics[scale=1.0]{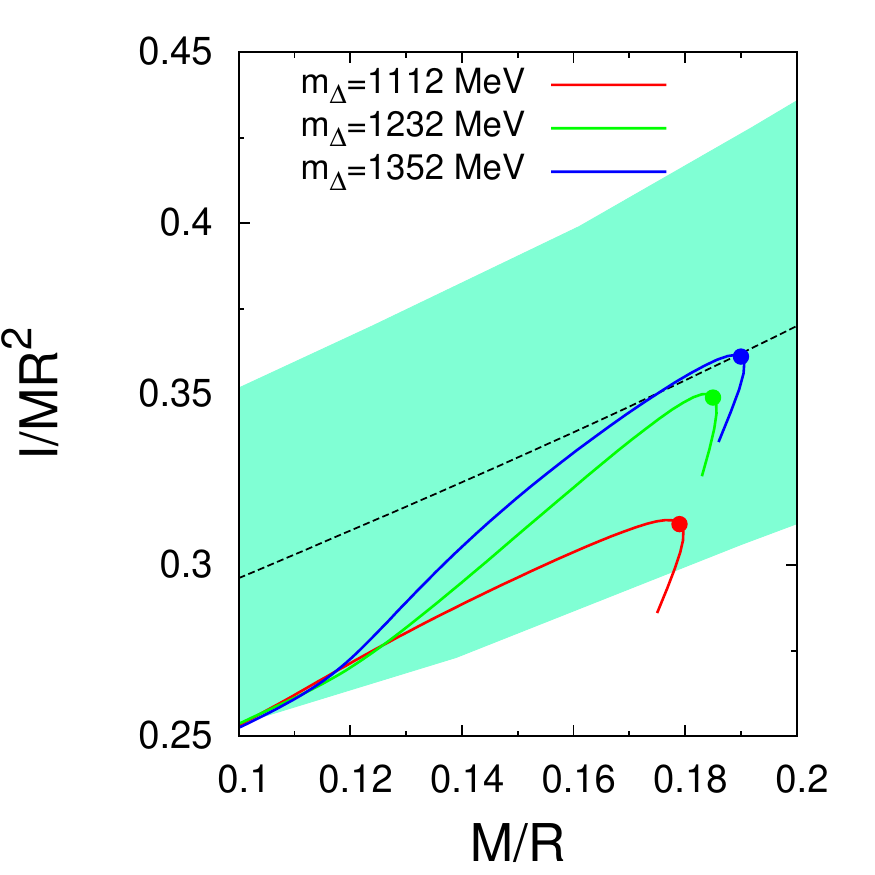}
\caption{\label{mI2norm300} Normalized moment of inertia ($I/MR^2$) versus compactness factor ($C=M/R$) of hybrid stars for different masses of $\Delta$ baryons rotating with frequency $\nu=300$ Hz. The fitted function of normalized $I$ from various theoretical models for slow rotation (black dashed line) \cite{Lat_Sch} is shown along with the uncertainty region (shaded region) \cite{Breu_Rez}}.
\end{figure}

\begin{figure}[!ht]
\centering
\includegraphics[scale=1.0]{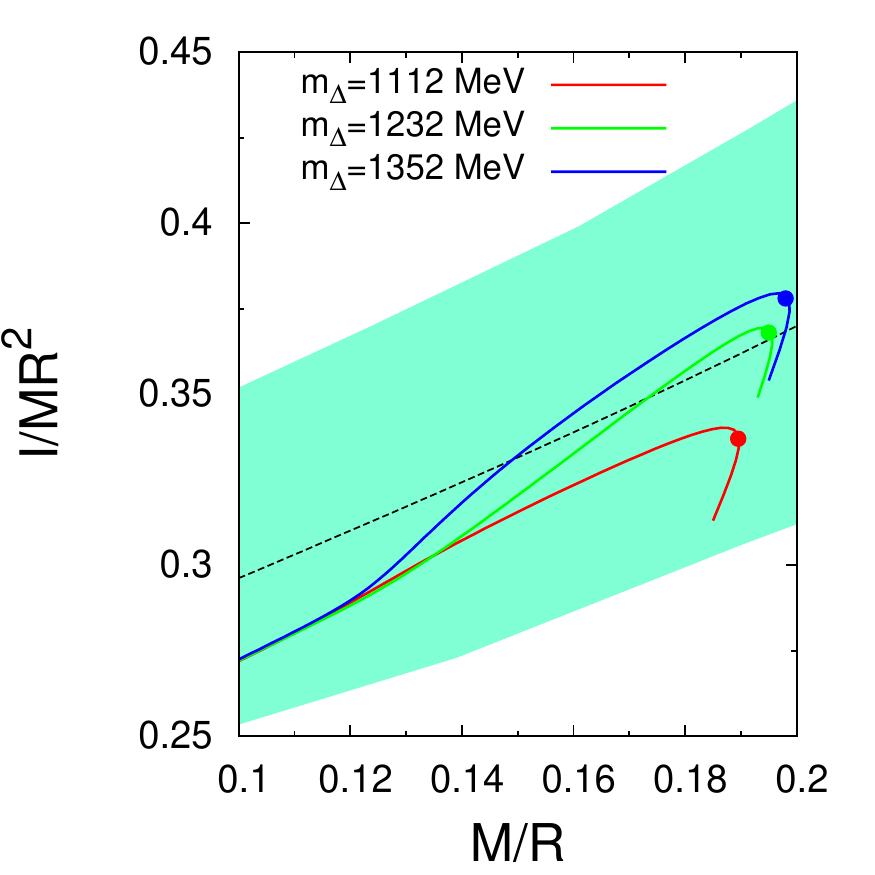}
\caption{\label{mI2norm600} Same as figure \ref{mI2norm300} but for $\nu=600$ Hz.}
\end{figure}

\begin{figure}[!ht]
\centering
\includegraphics[scale=0.9]{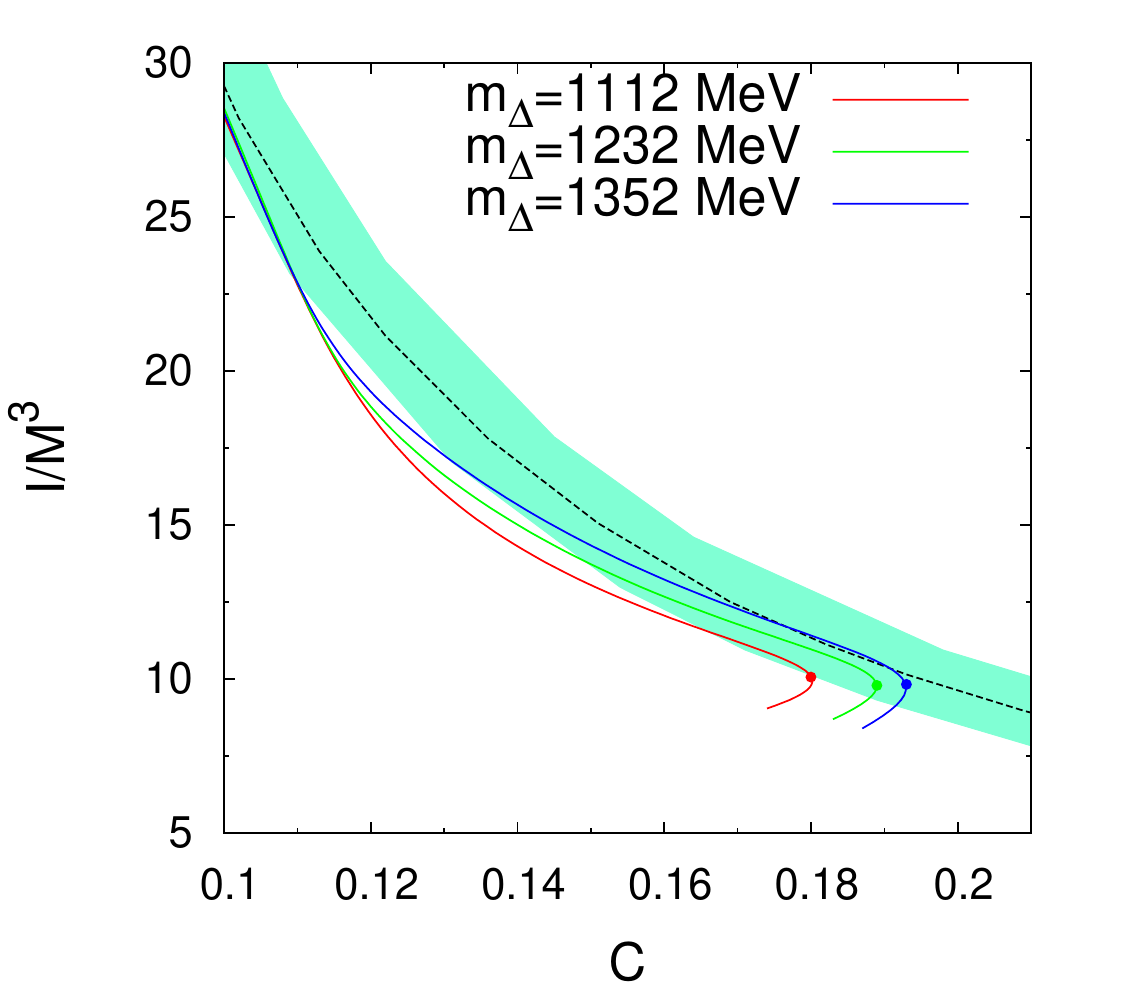}
\caption{\label{mInorm300} Normalized moment of inertia ($I/M^3$) versus compactness factor ($C=M/R$) of hybrid stars for different masses of $\Delta$ baryons rotating with frequency $\nu=300$ Hz.he fitted value of normalized I from various theoretical models for slow rotation (black dashed line) \cite{Lat_Sch} is shown along with the uncertainty region (shaded region) \cite{Breu_Rez}}.
\end{figure}

\begin{figure}[!ht]
\centering
\includegraphics[scale=0.9]{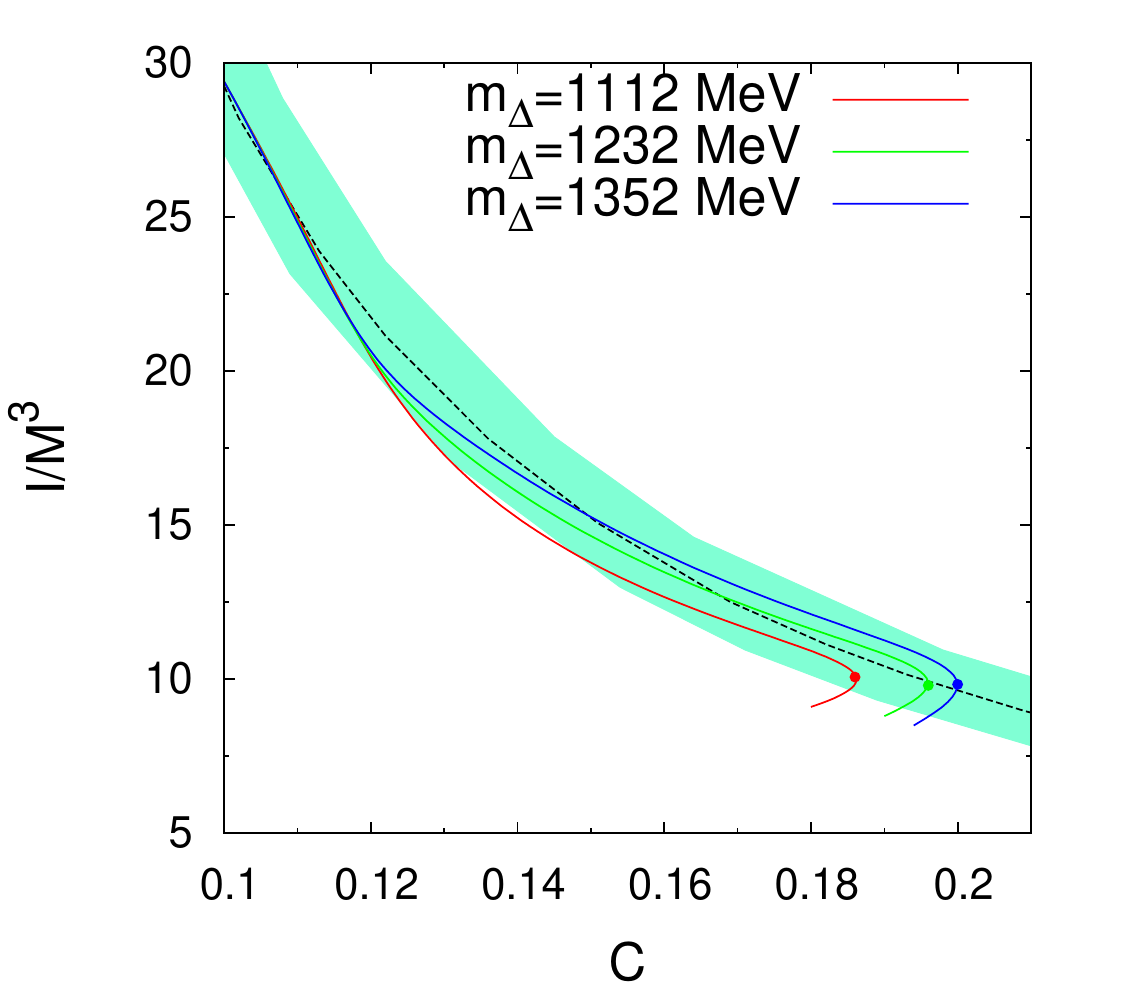}
\caption{\label{mInorm600} Same as figure \ref{mInorm300} but for $\nu=600$ Hz.}
\end{figure}

 In order to test the universality of the obtained hybrid EoS, the normalized moment of inertia $I/MR^2$ and $I/M^3$ are obtained for the three hybrid EoS considering slow rotation ($\nu=300$ and $\nu=600$ Hz). Figures \ref{mI2norm300} and \ref{mI2norm600} depict the variation of $I/MR^2$ with respect to the compactness parameter ($C=M/R$) for $\nu=300$ and $\nu=600$ Hz, respectively for all the hybrid EoS while figures \ref{mInorm300} and \ref{mInorm600} show the change of $I/M^3$ with $C$ for $\nu=300$ and $\nu=600$ Hz, respectively.

 It is clear from figures \ref{mI2norm300}, \ref{mI2norm600}, \ref{mInorm300} and \ref{mInorm600} that the universality of the hybrid EoS for all the values of $m_{\Delta}$ holds quite good in terms of normalized normalized moment of inertia. The obtained estimates of $I/MR^2$ and $I/M^3$ are in well agreement with the theoretical constraints from \cite{Breu_Rez} and \cite{Lat_Sch}. 

 The various rotational properties of HSs obtained from the present analysis are tabulated in table \ref{Rot.Prop}.

\begin{table}[ht!]
\caption{Rotational properties of hybrid stars for different masses of $\Delta$ baryons.}
\begin{center}
\begin{tabular}{cccccccccc}
\hline
\hline
\multicolumn{1}{c}{$m_{\Delta}$}&
\multicolumn{1}{c}{$\nu$}&
\multicolumn{1}{c}{$M$}&
\multicolumn{1}{c}{$M_{B}$} &
\multicolumn{1}{c}{$R$} &
\multicolumn{1}{c}{$I$} & \\
\multicolumn{1}{c}{(MeV)} &
\multicolumn{1}{c}{(Hz)} &
\multicolumn{1}{c}{($M_{\odot}$)} &
\multicolumn{1}{c}{($M_{\odot}$)} &
\multicolumn{1}{c}{($km$)} &
\multicolumn{1}{c}{($ 10^{45}$ g cm$^{2}$)} &\\
\hline
\hline
1112 &300      &2.08 &2.14 &11.8 &1.81 \\
     &600      &2.20 &2.29 &11.9 &2.10 \\
     &$\nu_K$  &2.60 &2.71 &12.8 &- \\
\hline     
1232 &300      &2.25 &3.10 &11.9 &2.26 \\
     &600      &2.39 &2.47 &12.2 &2.65 \\
     &$\nu_K$  &2.74 &2.83 &13.4 &- \\
\hline     
1352 &300      &2.40 &3.10 &12.3 &2.62 \\
     &600      &2.50 &3.20 &12.6 &3.05 \\
     &$\nu_K$  &2.89 &2.95 &14.1 &- \\
\hline
\hline
\end{tabular}
\end{center}

\protect\label{Rot.Prop}
\end{table}
  
 The present work shows that the $\Delta$ baryons play a significant role in determining the properties of HSs. In order to emphasize the effects of the uncertainty of $\Delta$ baryon mass on the HS properties, the latter has been calculated using a fixed value of bag constant $B$ and repulsive parameter $\alpha$ while $m_{\Delta}$ has been varied. A large number of properties of the HSs have been calculated in both static and rotating conditions and compared with the bounds obtained on such properties from different perspectives. One of the interesting results of this work is that the combined effect of $\Delta$s and phase transition shows large and interesting variation of $C_s^2$ values within the limit obtained from GW170817. Also the combined effect of formation of $\Delta$s and phase transition yields very compact HS configurations that helped to satisfy the radius constraints obtained from GW170817 better compared to that obtained with $\beta$ stable matter (fig. \ref{mr_all}). Also with the obtained HS configuration, the rotational properties are thoroughly examined and they are successfully consistent with the various constraints from different perspectives. Especially, the constraints on rotational frequency and moment of inertia from a wide variety of sources are compared and the results of the present work are found to be consistent with them.
  
\section{Summary and Conclusion}
\label{conclusion}

 The possibility of deconfinement of hadronic matter into quark matter at high density relevant to HS cores is studied in the present work. The effective chiral model is adopted to account for the hadronic matter while the MIT bag model describes the quark phase. The hadronic EoS is obtained by varying the mass of the $\Delta$ baryons in the presence of the hyperons. With Gibbs construction the hybrid EoS is obtained and the phase transition properties like the critical density of appearance of quarks, the density range for the persistence of the mixed phase, the population of different hadrons and quarks, the speed of sound in HSM are thoroughly studied. For the stiffest EoS, the quarks formation is delayed but the mixed phase persists the longest. For all the hybrid EoS, the maximum values of speed of sound is found to be within the bounds specified from GW170817 data analysis. 

 With considerably stiffened EoS due to phase transition, various static and rotational properties of the HS are calculated. In static case the gravitational mass estimates for $m_{\Delta}=1232~ \&~ 1352$ MeV are consistent with the bounds obtained from observational analysis of massive pulsars like PSR J0348+0432 and PSR J0740+6620. Also the radii estimates of $R_{1.4}$ and $R_{1.6}$ fall within the range suggested by analysis of GW170817 data from BNSM. Moreover, the static $M-R$ solutions are in excellent agreement with the NICER experimental data for PSR J0030+0451. With the hybrid EoS, the constraints on baryonic mass from PSR J0737-3039B and that on maximum surface redshift from EXO 07482-676, 1E 1207.4-5209 and RX J0720.4-3125 are also satisfied. With the formation of more $\Delta$ baryons, the constraint on baryonic mass from PSR J0737-3039B is also well satisfied.
 
 The rotational properties of HSs are also studied in the present work at different rotational frequencies. The bounds on maximum rotational frequency from fast rotating pulsars like PSR B1937+21, PSR J1748-2446ad and XTE J1739-285 are satisfied with all the HS configurations. The moment of inertia, studied for the slow rotation approximation ($\nu=300~\&~600$ Hz), satisfy the constraint from PSR J0737-3039A. With the intention to test the universality of the hybrid EoS, the dependence of normalized moment of inertia is studied with respect to compactness parameter and the universality holds quite good for all the HS configurations.
 
 Overall, the work presents a picture of the possible formation of various hadrons and quarks at relevant individual densities and consequently the gross structural properties of the resultant HSs are calculated. One of the primary aims of this work is to test the calculated structural properties of HSs with respect to the various recent constraints obtained from various perspectives. The work highlights that the uncertainty in $\Delta$ baryon mass may play an important role in satisfying these constraints. Additionally, it is also seen that reasonable HS configurations can also be obtained in static conditions with the variation of bag constant for a particular value of $\Delta$ baryon mass. The increasing values of bag constant yield massive and compact HS configurations upto a certain extent beyond which the solutions become unstable.

\appendix
\section{Effect of Bag constant on static hybrid star properties}
 
\begin{figure}[!ht]
\centering
\includegraphics[scale=1.2]{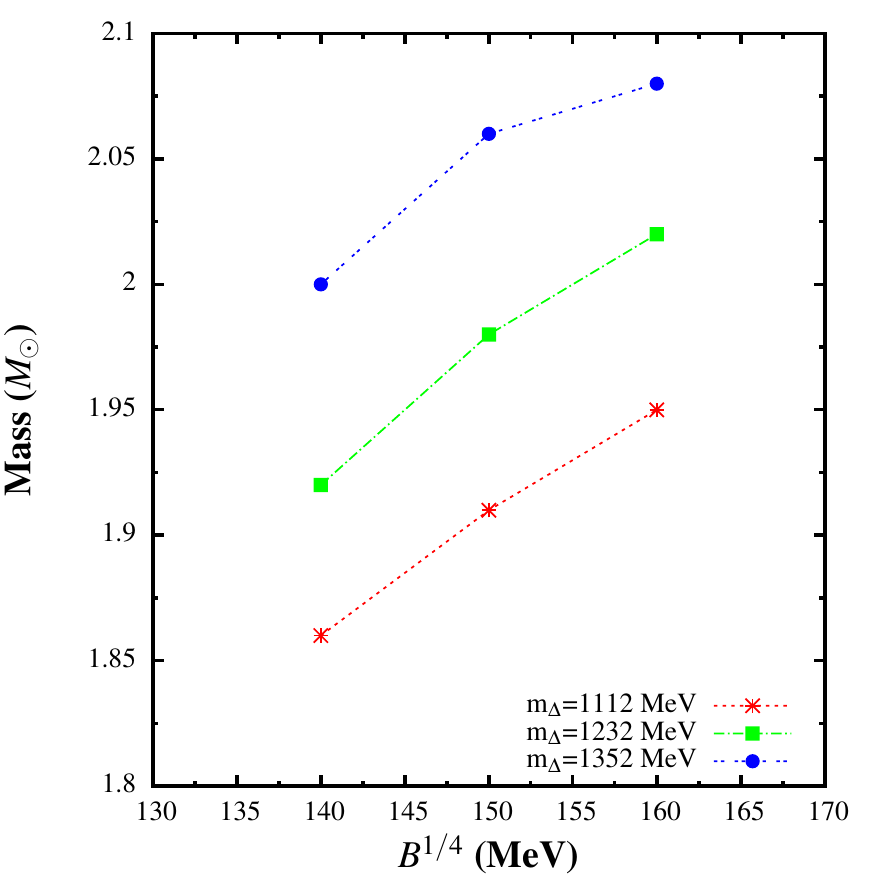}
\caption{\label{mB} Variation of maximum gravitational mass with Bag constant}.
\end{figure}

\begin{figure}[!ht]
\centering
\includegraphics[scale=1.2]{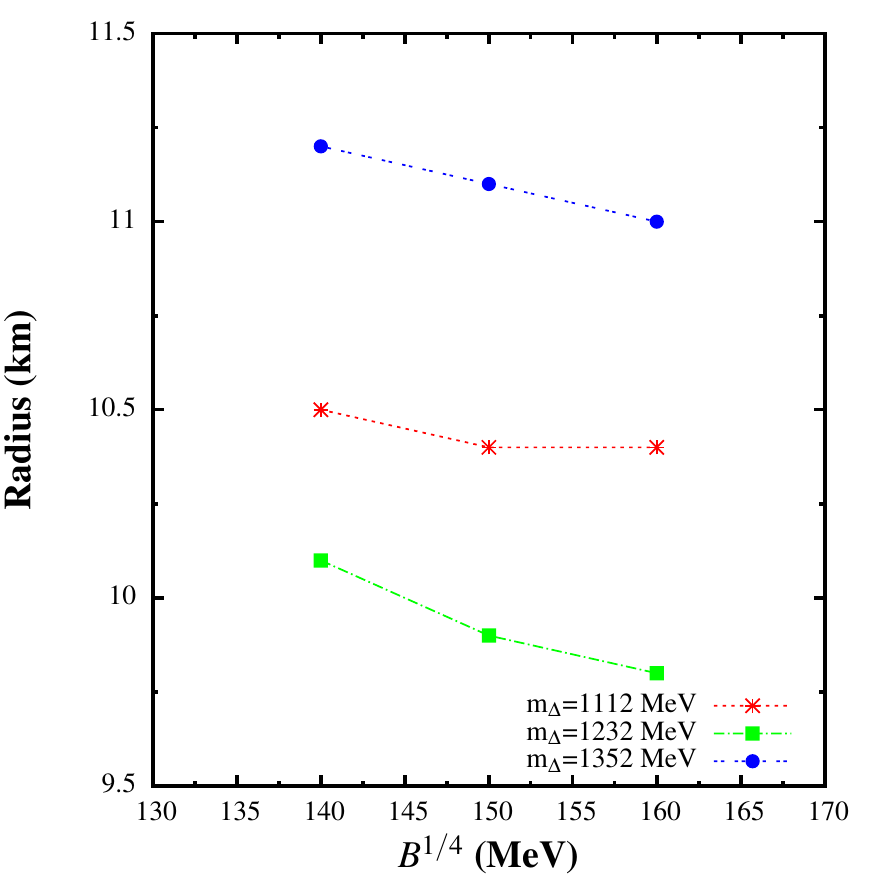}
\caption{\label{rB} Variation of radius with Bag constant}.
\end{figure}

 I now investigate the dependence of structural properties of HSs with respect to the bag constant in static condition. I show the variations of maximum mass (figure \ref{mB}) and the corresponding radius (figure \ref{rB}) with respect to the bag constant. The maximum limit to $B$ has been has been chosen consistent with that prescribed from \cite{Nandi,Rather} in the light of GW170817 data.
  
 Consistent with \cite{Ozel2010,Weissenborn2011,Klahn,Bonanno,Lastowiecki,Drago2016,Drago2016(2), Bombaci2016,Bombaci2017,Sen,Sen2}, I find that the increase in bag constant yields more massive and compact HS configurations. This is because for a given hadronic EoS obtained with a fixed value of $m_{\Delta}$, the higher value of bag constant shifts the hadron-quark crossover points to higher densities and the delayed transition results in stiffer hybrid EoS and consequently massive HS configurations. However, I find that a too delayed transition leads to an unstable solution of HS configuration as in the last possibility ($m_{\Delta}$=1352 MeV; $B^{1/4}$=160 MeV). For $m_{\Delta}$=1112 MeV, the increase in maximum mass is upto 4.8\% while for $m_{\Delta}$=1232 MeV and 1352 MeV it is 5.2\% and 4\%, respectively. The radius shows slight decrease in value with the increase in bag constant. The values of maximum mass and corresponding radius of HSs obtained by varying $B^{1/4}$ for fixed values of $m_{\Delta}$ are tabulated in table \ref{BMR}.

\begin{table}[ht!]
\caption{Variation of maximum mass and corresponding radius of hybrid stars with Bag constant for different masses of $\Delta$ baryons.}
\begin{center}
\begin{tabular}{cccccc}
\hline
\hline
\multicolumn{1}{c}{$m_{\Delta}$}&
\multicolumn{1}{c}{$B^{1/4}$}&
\multicolumn{1}{c}{$M$}&
\multicolumn{1}{c}{$R$} &\\
\multicolumn{1}{c}{(MeV)} &
\multicolumn{1}{c}{(MeV)} &
\multicolumn{1}{c}{($M_{\odot}$)} &
\multicolumn{1}{c}{($km$)} &\\
\hline
\hline
1112 &140  &1.86 &10.5  \\
     &150  &1.91 &10.4  \\
     &160  &1.95 &10.4  \\
\hline     
1232 &140  &1.92 &10.1  \\
     &150  &1.98 &9.9  \\
     &160  &2.02 &9.8  \\
\hline     
1352 &140  &2.00 &11.2  \\
     &150  &2.06 &11.1  \\
     &160  &2.08 &11.0  \\
\hline
\hline
\end{tabular}
\end{center}
\protect\label{BMR}
\end{table}

\end{document}